\documentclass[twocolumn,aps,showpacs,superscriptaddress]{revtex4-1}
\usepackage{amsmath}
\usepackage{graphicx}
\RequirePackage[pagewise,mathlines]{lineno}

\begin{document}
%\linenumbers
\title{Coherent J/$\psi$ photoproduction in hadronic heavy-ion collisions}%
\author{W. Zha}\affiliation{University of Science and Technology of China, Hefei, China}
\author{S.R. Klein}\affiliation{Lawrence Berkeley National Laboratory, Berkeley, California, USA}
\author{R. Ma}\affiliation{Brookhaven National Laboratory, New York, USA}
\author{L. Ruan}\affiliation{Brookhaven National Laboratory, New York, USA}
\author{T. Todoroki}\affiliation{Brookhaven National Laboratory, New York, USA}
\author{Z. Tang}\email{zbtang@ustc.edu.cn}\affiliation{University of Science and Technology of China, Hefei, China}
\author{Z. Xu}\affiliation{Brookhaven National Laboratory, New York, USA}\affiliation{Shandong University, Jinan, China}
\author{C. Yang}\affiliation{Shandong University, Jinan, China}
\author{Q. Yang}\affiliation{University of Science and Technology of China, Hefei, China}
\author{S. Yang}\affiliation{Brookhaven National Laboratory, New York, USA}
%\author{The American Physical Society}%
%\email[REVTeX Support: ]{revtex@aps.org}
%\affiliation{1 Research Road, Ridge, NY 11961}
\date{\today}%
\begin{abstract}
Significant excesses of J/$\psi$ yield at very low transverse momentum ($p_T < 0.3$ GeV/c) were observed by the ALICE and STAR collaborations in peripheral hadronic A+A collisions. This is a sign of coherent photoproduction of J/$\psi$ in violent hadronic interactions. Theoretically, the photoproduction of J$/\psi$ in hadronic collisions raises questions about how spectator and non-spectator nucleons participate in the coherent reaction. We argue that the strong interactions in the overlapping region of incoming nuclei may disturb the coherent production, leaving room for different coupling assumptions. The destructive interference between photoproduction on ions moving in opposite directions also needs to be included.
% The survival of photoproduced J/$\psi$ merits theoretical investigation.
%In this letter we report on

This letter presents calculations of J$/\psi$ production from coherent photon-nucleus ($\gamma + A \rightarrow \text{J}/\psi + A$) interactions in hadronic A+A collisions at RHIC and LHC energies with both nucleus and spectator coupling hypotheses. The integrated yield of coherent J/$\psi$ as a function of centrality is found to be significantly different, especially towards central collisions, for different coupling scenarios. Differential distributions as a function of transverse momentum, azimuthal angle and rapidity in different centrality bins are also shown, and found to be more sensitive to the Pomeron coupling than to the photon coupling. These predictions call for future experimental measurements to help better understand the coherent interaction in hadronic heavy-ion collisions.  %We
%address the questions about
%discuss how the electromagnetic field translates into a flux of equivalent photons in hadronic A+A collisions: if the photons are emitted from the whole nucleus, or if only the spectator fragments contribute to the photon emission. Similarly, both the whole nucleus and spectator fragment are considered as the photon target. Model calculations with different scenarios for the photon emitter and the target are compared with the experimental results, and further differential measurements are proposed to distinguish different scenarios.
\end{abstract}
\maketitle
%\tableofcontents
In relativistic heavy-ion collisions carried out at the Relativistic Heavy Ion Collider (RHIC) and the Large Hadron Collider (LHC), one aims at searching for a new form of matter --- the Quark-Gluon Plasma (QGP)~\cite{PBM_QGP} and studying its properties in laboratory. J/$\psi$ suppression, due to the color screening effect in the deconfined medium, was proposed as a direct signature of the QGP formation~\cite{MATSUI1986416}. Other mechanisms, such as the recombination of deconfined charm quarks in the QGP and cold nuclear matter (CNM) effects, also play a significant role in affecting the J/$\psi$ yield. The interplay of these effects can qualitatively describe the J/$\psi$ production measured so far at SPS, RHIC and LHC~\cite{STAR_Jpsi_AuAu_BES}.

J/$\psi$ can also be produced via the coherent photon-nucleus interactions through the photon-Pomeron fusion in heavy-ion collisions~\cite{UPCreview}. Virtual  photons emitted by one nucleus may fluctuate into $q\overline q$ pairs, scatter off the other nucleus and emerge as vector mesons.  The coherent nature of the interactions leads to a distinctive signature: the final products consist of two intact nuclei, a J/$\psi$ with very low transverse momentum ($p_{T} < 0.1$ GeV/c) and nothing else.
Conventionally, these reactions are only visible when they are not accompanied by hadronic interactions, $i.e.$ in the so-called Ultra-Peripheral Collisions (UPCs). In these collisions, the impact parameter ($b$) is larger than twice the nuclear radius ($R_A$).  %Coherent photon-Pomeron fusion is sensitive to the gluon distribution in the nucleus~\cite{Rebyakova2012647}, for which currently there exists considerable uncertainties in the low Bjorken-$x$ region~\cite{1126-6708-2009-04-065}.

Can the coherent photonuclear interaction also occur in Hadronic Heavy-Ion Collisions (HHICs, $b<2R_A$), where the nuclei collide and break up?
Recently, significant excesses of J/$\psi$ yield at very low $p_{T}$ ($<$ 0.3 GeV/c) have been observed by the ALICE~\cite{LOW_ALICE} and STAR~\cite{1742-6596-779-1-012039} collaborations in peripheral HHICs.  These excesses cannot be explained by the hadronic J/$\psi$ production with currently known cold and hot medium effects taken into account. Interestingly, the excesses exhibit characteristics of coherent photonuclear interactions. Klusek-Gawenda and Szczurek considered this problem in \cite{Klusek-Gawenda:2015hja}, where they assumed that both the photons and the photon targets couple to the whole nucleus and modify the photon flux by ignoring the overlapping region. However, the modification to the photon flux is not unambiguous, and furthermore, the destructive interference between photoproduction on ions moving in opposite directions was not included in their model.
%. The observed excess may originate from the coherent photon-nucleus interactions, which would be very challenging for the current understanding of coherent photon-Pomeron fusion.

In this letter, we consider different coupling scenarios for photons and Pomerons with the nucleus in presence of hadronic interactions. The interference between the J/$\psi$ photoproduction amplitudes on ions moving in opposite direction (INT2N) is also addressed, which turns out to play a more significant role in HHICs than that in UPCs.  %that break up the nuclei and even produce QGP.
The coherent J/$\psi$ yields in HHICs are calculated at RHIC and LHC energies and compared with experimental results. Furthermore, differential distributions as a function of transverse momentum, centrality and rapidity are also shown.

The cross-section for J/$\psi$ production via the photon-Pomeron fusion can be calculated by convoluting the Weizs\"acker-Williams virtual photon spectrum with the photonuclear interaction cross-section~\cite{UPC_JPSI_PRC,UPC_JPSI_PRL}:
  \begin{equation}
  \label{equation1}
  \sigma(\text{AA} \rightarrow \text{AA} \text{J}/\psi) = \int d\omega_{\gamma}\frac{dN_{\gamma}(\omega_{\gamma})}{d\omega_{\gamma}}\sigma(\gamma A \rightarrow \text{J}/\psi A)
  \end{equation}
where $\omega_{\gamma}$ is the photon energy, and $\sigma(\gamma A \rightarrow \text{J}/\psi A)$ is the photonuclear interaction cross-section for J$/\psi$. It is determined from measurements of $\gamma p$ interactions coupled with a Glauber formalism~\cite{Klein:2016yzr,PhysRevC.67.034901}. In UPCs where photons and Pomerons couple coherently to the entire nucleus, the induced photon flux is given by the equivalent photon approximation~\cite{KRAUSS1997503}:
  \begin{equation}
  \label{equation2}
  \begin{aligned}
  & \frac{d^{3}N_{\gamma}(\omega_{\gamma},\vec{x}_{\bot})}{d\omega_{\gamma} d\vec{x}_{\bot}} = \frac{4Z^{2}\alpha}{\omega_{\gamma}}\bigg|\int\frac{d^{2}\vec{k}_{\gamma\bot}}{(2\pi)^{2}}\vec{k}_{\gamma\bot}\frac{F_{\gamma}(\vec{k}_{\gamma})}{|\vec{k}_{\gamma}|^{2}}
  e^{i\vec{x}_{\bot}\cdot\vec{k}_{\gamma\bot}}\bigg|^{2}
  \\
  & \vec{k}_{\gamma}=(\vec{k}_{\gamma\bot},\frac{\omega_{\gamma}}{\gamma_{c}}) \ \ \ \ \  \omega_{\gamma}=\frac{1}{2}M_{\text{J}/\psi} e^{\pm y}
  \end{aligned}
  \end{equation}
 where $\vec{x}_{\bot}$ and $\vec{k}_{\gamma\bot}$ are 2-dimensional photon position and momentum vectors perpendicular to the beam direction, $Z$ the nuclear charge, $\alpha$ the electromagnetic coupling constant, $\gamma_{c}$ the Lorentz factor of the photon-emitting nucleus, $M_{\text{J}/\psi}$ and $y$ the mass and rapidity of J/$\psi$, and $F_{\gamma}(\vec{k}_{\gamma})$ the nuclear electromagnetic form factor.  $F_{\gamma}(\vec{k}_{\gamma})$ is obtained via the Fourier transformation of the charge density in the nucleus. The charge density for a symmetrical nucleus $A$ is given by the Woods-Saxon distribution:
  \begin{equation}
  \rho_{A}(r)=\frac{\rho^{0}}{1+\exp[(r-R_{\rm{WS}})/d]}
  \label{equation2_new}
  \end{equation}
  where the radius $R_{\rm{WS}}$ and skin depth $d$ are based on fits to electron scattering data~\cite{0031-9112-29-7-028}, and $\rho^{0}$ is the normalization factor.
 The cross-section for the process $\gamma A \rightarrow \text{J}/\psi A$ can be derived from the following sequence of equations~\cite{Klusek-Gawenda:2015hja,UPC_JPSI_PRC}:
    \begin{equation}
    \label{equation3_1}
    \begin{split}
    &\sigma(\gamma A \rightarrow \text{J}/\psi A)=\frac{d\sigma(\gamma A \rightarrow \text{J}/\psi A)}{dt}\bigg|_{t=0} \times\\
     &\int|F_{P}(\vec{k}_{P})|^{2}d^{2}{\vec{k}_{P\bot}} \ \ \ \ \ \ \vec{k}_{P}=(\vec{k}_{P\bot},\frac{ \omega_{P}}{\gamma_{c}})\\
  & \omega_{P} = \frac{1}{2}M_{\text{J}/\psi} e^{\pm y} = \frac{M_{\text{J}/\psi}^{2}}{4\omega_{\gamma}}
    \end{split}
    \end{equation}
    \begin{equation}
    \label{equation3_2}
    \frac{d\sigma(\gamma A \rightarrow \text{J}/\psi A)}{dt}\bigg|_{t=0}=C^{2}\frac{\alpha \sigma_{tot}^{2}(\text{J}/\psi A)}{4f_{\text{J}/\psi}^{2}}
    \end{equation}
    \begin{equation}
    \label{equation3_3}
    \sigma_{tot}(\text{J}/\psi A)=2\int(1-\exp(-\frac{1}{2}\sigma_{tot}(\text{J}/\psi p)T_{A}(x_{\bot})))d^{2}x_{\bot}
    \end{equation}
    \begin{equation}
    \label{equation3_4}
    \sigma_{tot}^{2}(\text{J}/\psi p)=16\pi\frac{d\sigma(\text{J}/\psi p \rightarrow \text{J}/\psi p)}{dt}\bigg|_{t=0}
    \end{equation}
    \begin{equation}
    \label{equation3_5}
    \frac{d\sigma(\text{J}/\psi p \rightarrow \text{J}/\psi p)}{dt}\bigg|_{t=0}=\frac{f_{\text{J}/\psi}^{2}}{4\pi \alpha C^{2}}\frac{d\sigma(\gamma p \rightarrow \text{J}/\psi p)}{dt}\bigg|_{t=0}
    \end{equation}
where $T_{A}(x_{\bot})$ is the nuclear thickness function, $-t$ is the squared four momentum transfer, $f_{\text{J}/\psi}$ is the J/$\psi$-photon coupling, $\omega_{P}$ is the energy of Pomeron and $C$ is a correction factor, which will be discussed in detail hereinafter. Parametrization for $\gamma p \rightarrow \text{J}/\psi p$ production in Eq.~\ref{equation3_5} is obtained from~\cite{Klein:2016yzr}.

As shown in the equation sequences (Eq.~\ref{equation3_1} - Eq.~\ref{equation3_5}), the calculation of $\sigma(\gamma A \rightarrow \text{J}/\psi A)$ are performed with a quantum Glauber approach coupled with the parameterized $\sigma(\gamma p \rightarrow \text{J}/\psi p)$ as input. To relate this to nuclei, we follow vector dominance model~\cite{RevModPhys.50.261} and make use of the optical theorem and an Eikonalization technique. However, the single vector dominance model failed to describe the $\gamma p \rightarrow \text{J}/\psi p$ cross section compared to the absorption J/$\psi$ cross section extracted from nuclear data~\cite{HUFNER1998154}. A correction is required to account for the non-diagonal coupling through higher mass vector mesons, as implemented in the generalized vector dominance model~\cite{HUFNER1998154,PhysRevC.57.2648}. For J/$\psi$, the correction factor $C$ derived in Ref.~\cite{HUFNER1998154} ($C = 0.3$) is adopted in our calculation. Since the same correction factor is used both in Eq.~\ref{equation3_2} and Eq.~\ref{equation3_5}, its effect largely cancels, so does the uncertainty associated with it. The $F_{P}(\vec{k}_{P})$ in Eq.~\ref{equation3_1} is the nuclear form factor for Pomeron and can be obtained by performing a Fourier transformation of the nuclear density in the nucleus, which is assumed to be the same as the charge density of the nucleus. Here $\vec{k}_{P} = (\vec{k}_{P\bot},\vec{k}_{L})$, the longitudinal component $k_{L}=\frac{M_{\text{J}/\psi}^{2}}{4\gamma_{c}\omega_{\gamma}}$ is the momentum transfer required to produce a real J/$\psi$. It leads to a coherence length of $\hbar/k_{L}$. As long as $\hbar/k_{L}$ is larger than twice the nuclear radius, the reaction is fully longitudinally coherent. Any longitudinal destructive interference has been taken into account via the phase factor $e^{ik_{L}z}$ in the Pomeron form factor.

We now extend these calculations to HHICs, where, unlike in the UPCs, the incoming nuclei collide and break up. In UPCs, coherent $\rho^0$ photoproduction is seen to be unaffected by the accompanying mutual Coulomb excitation~\cite{Baltz:2002pp,Abelev:2007nb}. This can be attributed to the long lifetime of the excited nuclei compared to the coherent emission of photons and Pomerons. However, in HHICs, the more energetic hadronic interactions happen at a much smaller time scale, and therefore could impose significant impact on the coherent photoproduction. This possible disruptive effect is considered for two distinct sub-processes: photon emission and Pomeron emission.

For photon emission, the photon field travels along with the incoming nucleus and arrives at the target at the same time as the emitter. Since the photons are nearly real, $i.e.$ $Q^2  < (\hbar/R_A)^2$, they are likely to be emitted before the hadronic interactions occur by about $\Delta t = R_A/c$. Therefore the photon emission should be unaffected by hadronic interactions. However, one needs to take into account the transverse extent of the photon emitter as the two colliding nuclei overlap. For example, the nucleons located in the overlapping region of the target nucleus should see a reduced photon flux since the effective photon flux decreases rapidly towards the center of the emitter. In fact, the photon flux vanishes at the center of the emitting nucleus by symmetry. Given that, two limiting cases are considered for the photon emission, $i.e.$ either the entire nucleus or only the spectator nucleons act as the emitter.

 For Pomeron emission, the spectator nucleons, which are free from the hadronic interactions, can still act coherently. On the other hand, for the participating nucleons, the state is likely to be affected by the violent hadronic interactions, leading to the destruction of coherent Pomeron emission. In addition, the losses of longitudinal momenta for nucleons involving in hadronic interactions are significant, leading also to a decease of photoproduction cross-section. Finally, to determine whether the participating nucleons act coherently, one needs to examine the time ordering of the hadronic interaction and the coherent process. These interactions can be ordered in terms of the formation time, $i.e.$ $\hbar/M_{\text{J}/\psi}$ for the coherent J/$\psi$ production and $R_{A}/\gamma c$ for the hadronic interactions, which turn out to be of the same order. To make things even more complicated, time ordering is not Lorentz invariant, which means the Feynman diagrams of all possible time orderings need to be summed up in order to obtain the correct cross-section~\cite{Drell:1970yt}. Since a full solution to the time ordering problem is currently unavailable, two limiting scenarios for the Pomeron emission are considered as well. The first is to ignore the hadronic interactions, and assume that the entire nucleus acts coherently in emitting Pomerons. The second is to take only the spectator nucleons as the coherent emitter. These two scenarios should bracket the actual case.

In the end, four different coupling scenarios are considered for the coherent J/$\psi$ production: (1) Nucleus (photon emitter) + Nucleus (Pomeron emitter), short for ``N+N'' here; (2) Nucleus + Spectator (``N+S''); (3) Spectator + Nucleus (``S+N'') and (4) Spectator + Spectator (``S+S''). The collision geometry and the density of spectators are simulated by the optical Glauber model~\cite{doi:10.1146/annurev.nucl.57.090506.123020}.

 The transverse momentum of coherently produced J$/\psi$ is equal to the sum of the perpendicular momenta ($\vec{k}_{\bot}$) of the incoming photon and Pomeron~\cite{UPC_PT}. The photon perpendicular momentum ($\vec{k}_{\gamma\bot}$) spectrum is given by the equivalent photon approximation \cite{KRAUSS1997503}:
  \begin{equation}
  \label{equation4}
  \frac{d^{2}N_{\gamma}}{d^{2}\vec{k}_{\gamma\bot}} = K_{0}\frac{F_{\gamma}^{2}(\vec{k}_{\gamma})\vec{k}^{2}_{\gamma\bot}}{(\vec{k}_{\gamma\bot}^{2}+\omega^{2}_{\gamma}/\gamma_{c}^{2})^{2}}
  \end{equation}
  where $K_{0}$ is the dimensionless normalization factor.
The Pomeron  perpendicular momentum ($\vec{k}_{P\bot}$) spectrum is given by the nuclear form factor of the emitter:
   \begin{equation}
  \label{equation5}
  \begin{split}
  &\frac{d^{2}N_{P}}{d^{2}\vec{k}_{P\bot}} = N_{0}F_{P}^{2}(\vec{k}_{P})
  \end{split}
  \end{equation}
where $N_{0}$ is the normalization factor with dimension $\rm{GeV^{-2}}$.

For J$/\psi$ with $p_{T} < \hbar /b$, it is impossible to distinguish which nucleus emits the photon, and which emits the Pomeron. Due to the negative parity of J/$\psi$, the signs of the two amplitudes are opposite, leading to destructive interference. This INT2N effect has been studied in detail by Klein and Nystrand~\cite{UPC_PT} for the vector meson production in UPCs, and verified by the STAR measurements of coherent $\rho^{0}$ production \cite{Abelev:2008ew}. We follow the same strategy as in~\cite{UPC_PT} for coherent J/$\psi$ production:
    \begin{equation}
    \begin{aligned}
    &\sigma(p_{T},y,b) = A^{2}(p_{T},y,b) +A^{2}(p_{T},-y,b)
    \\
    & -2A(p_{T},y,b)A(p_{T},-y,b) \times cos(\vec{p}_{T} \cdot \vec{b})
         \label{equation6}
    \end{aligned}
    \end{equation}
where $A(y,p_{T},b)$ is the amplitude for J/$\psi$ production at rapidity y with transverse momentum $p_{T}$. Unlike in the UPCs, the impact parameters of HHICs can be related to the collision centrality, usually determined experimentally by measuring event activities in certain rapidity ranges, using the Glauber model~\cite{Loizides:2014vua}. This makes it possible to compare the measured $p_T$ spectra of coherent J/$\psi$ production in different centrality classes to the theoretical calculations for corresponding impact parameter ranges to study the INT2N effect differentially.
\renewcommand{\floatpagefraction}{0.75}
\begin{figure}[htbp]
\includegraphics[keepaspectratio,width=0.5\textwidth]{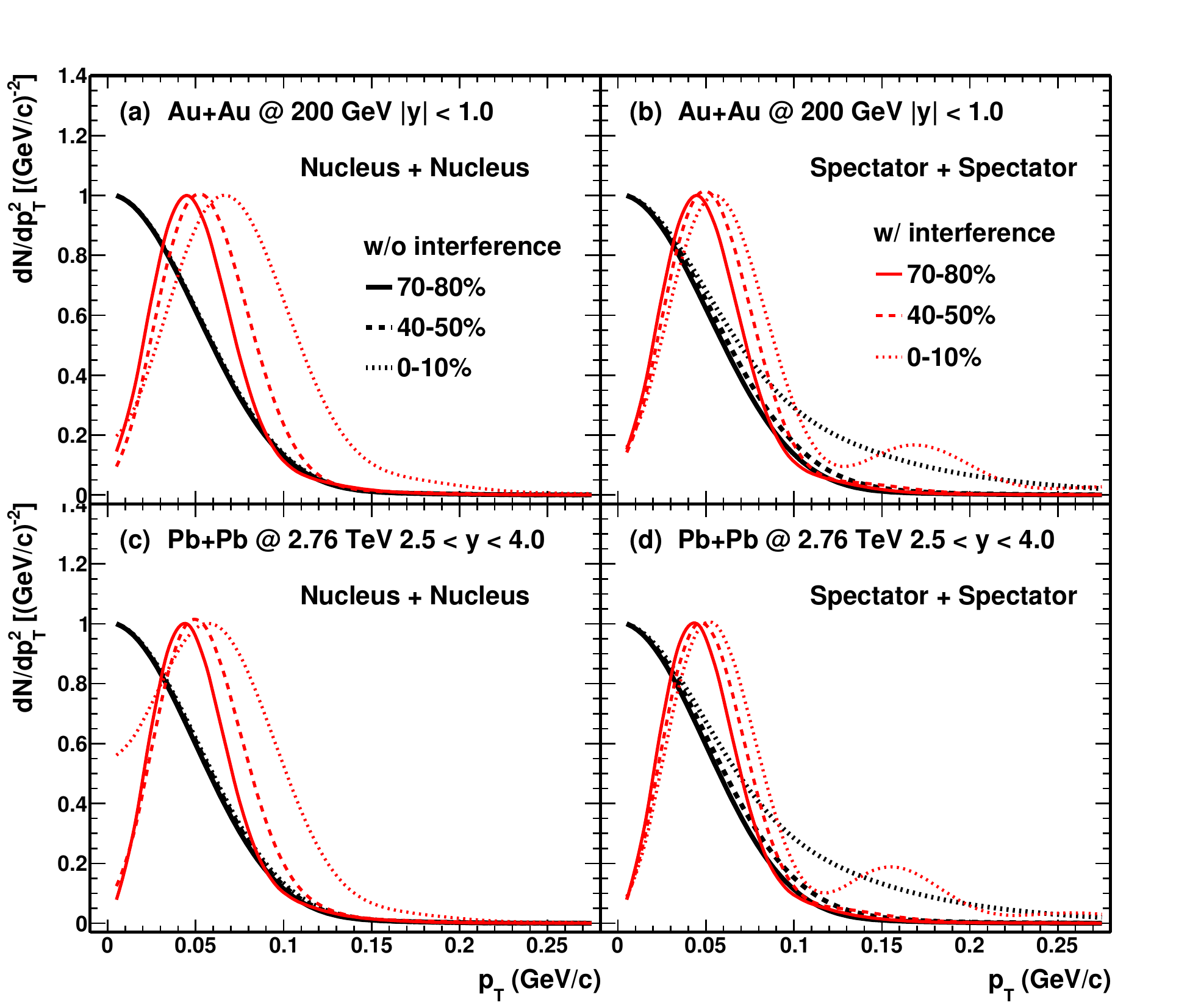}
\caption{The $dN/dp_{T}^{2}$ distributions of coherent J/$\psi$ for different centrality classes with the ``N+N'' (panel a and c) and the ``S+S'' (panel b and d) scenarios. The top two panels are for Au+Au collisions at $\sqrt{s_{\text{NN}}}$ = 200 GeV at mid-rapidity ($|y| < 1$), while the bottom two for Pb+Pb collisions at $\sqrt{s_{\text{NN}}}$ = 2.76 TeV at forward rapidity ($2.5<y<4.0$). The black curves represent the calculations without INT2N, while the red ones with peak structures denote results with INT2N. All the distributions are normalized such that the maximum values are equal to 1.}
\label{figure4}
\end{figure}

Figure \ref{figure4} shows the $dN/dp_{T}^{2}$ distributions for coherent J/$\psi$ in different centrality bins with the ``N+N'' (panel a and c) and ``S+S'' (panel b and d) scenarios. The top two panels show the predictions for Au+Au collisions at $\sqrt{s_{\text{NN}}}$ = 200 GeV at mid-rapidity ($|y| < 1$), while the bottom two are for Pb+Pb collisions at $\sqrt{s_{\text{NN}}}$ = 2.76 TeV at forward rapidity ($2.5<y<4.0$). The red lines with peak structures include INT2N, while the black ones do not. Without the INT2N effect, the shapes of the coherent J/$\psi$ $p_T$ spectra show negligible dependence on the collision centrality when both the photon and the Pomeron couple to the entire nucleus. However, for the ``S+S'' scenario, sizable differences show up due to the different density profiles of the spectators, and the differences grow larger towards more central collisions. On the other hand, when the INT2N effect is included, a significant suppression of the coherent J/$\psi$ production at very low $p_T$ is seen, as expected. As the impact parameter gets smaller, the INT2N effect affects larger kinematic ranges, resulting in broader distributions and higher values of $\langle p_T \rangle$. Comparing different scenarios, the INT2N effect is more significant for the nucleus coupling than for the spectator coupling, due probably to the smaller distance between the two nuclei than that between the centroid of the spectator fragments. For the ``S+S'' scenario, benefiting from the relatively large production rate above 0.1 GeV/c in 0-10$\%$ central collisions, the second bump originating from the INT2N becomes visible. The shapes of ``S+N'' and ``N+S'' scenarios are very close to those of the ``N+N'' and ``S+S'' scenarios, respectively. This indicates that the coherent J/$\psi$ $p_T$ spectrum shape is more sensitive to the Pomeron emitter rather than the photon emitter.
%.  The $p_{T}$ spectra of J/$\psi$ are more sensitive to the target nucleus.
%Taken the interference effect into account, as depicted in the figure,
%it leads to significant suppression of
%J/$\psi$ production at low $p_{T}$.
%Because  $|b|$ decreases in more central collisions, the interference becomes more significant with increasing centrality, pushing the $\langle p_{T} \rangle$ of J/$\psi$ to higher values.
%Benefit from the relative large production at high $p_{T}$ in 0-10$\%$ central collisions with ``S+S'' scenario, the second bump originated from interference is significant in this case.   As expected, %when the Pomeron couple with nuclei, the interference is more significant than the case of spectator coupling, because the distance between the two nuclei is smaller than that between the spectators.
\renewcommand{\floatpagefraction}{0.75}
\begin{figure}[htbp]
\includegraphics[keepaspectratio,width=0.49\textwidth]{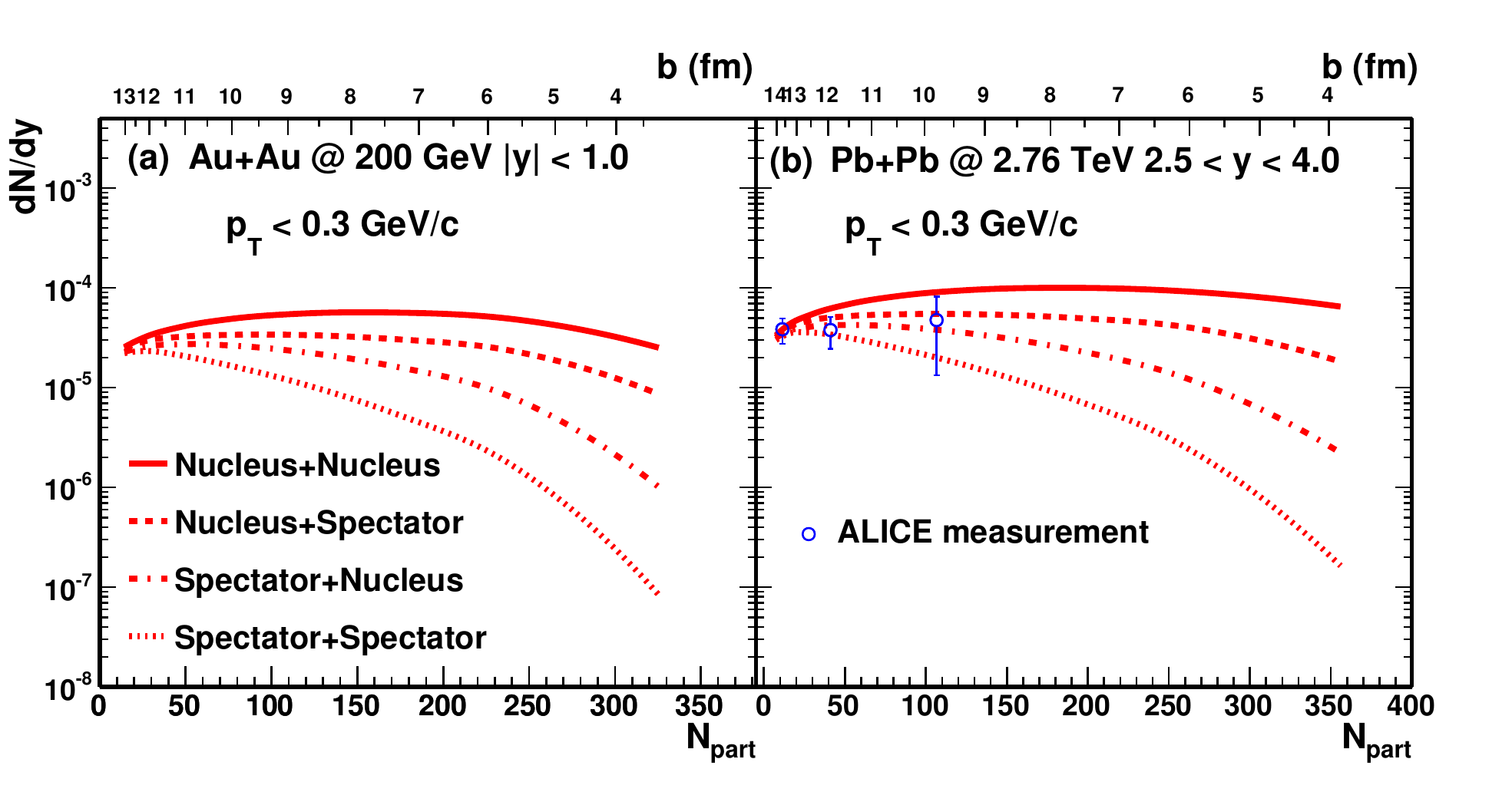}
\caption{Yields of coherent J/$\psi$ production as a function of $N_{\text{part}}$ in Au+Au collisions at $\sqrt{s_{\text{NN}}}$ = 200 GeV (a) and Pb+Pb collisions at $\sqrt{s_{\text{NN}}}$ = 2.76 TeV (b). Data from the ALICE experiment~\cite{LOW_ALICE} are shown for comparison.}
\label{figure1}
\end{figure}

Figure \ref{figure1} shows the coherent J/$\psi$ yield, including INT2N effects, as a function of number of participants ($N_{\text{part}}$) in Au+Au collisions at $\sqrt{s_{\text{NN}}}$ = 200 GeV (panel a) and Pb+Pb collisions at $\sqrt{s_{\text{NN}}}$ = 2.76 TeV (panel b).
%The different lines correspond to different scenarios as described in the legend of the figure.
The four scenarios, shown with different styles of lines in the figure, predict similar yields at $b=2R_A$, but differ dramatically as $b$ decreases. The ALICE data~\cite{LOW_ALICE} are consistent with all four scenarios within the uncertainties. Current calculations do not account for the nuclear shadowing effect on parton distribution functions (nPDFs). At the LHC, the measurements in UPCs show that the shadowing effect could reduce the cross-section significantly~\cite{Klein:2017vua}, while the effect is expected to be smaller at RHIC energies. Measurements of better precision towards more central collisions and advanced models with nPDFs included are essential for distinguishing the different scenarios.
\renewcommand{\floatpagefraction}{0.75}
\begin{figure}[htbp]
\includegraphics[keepaspectratio,width=0.5\textwidth]{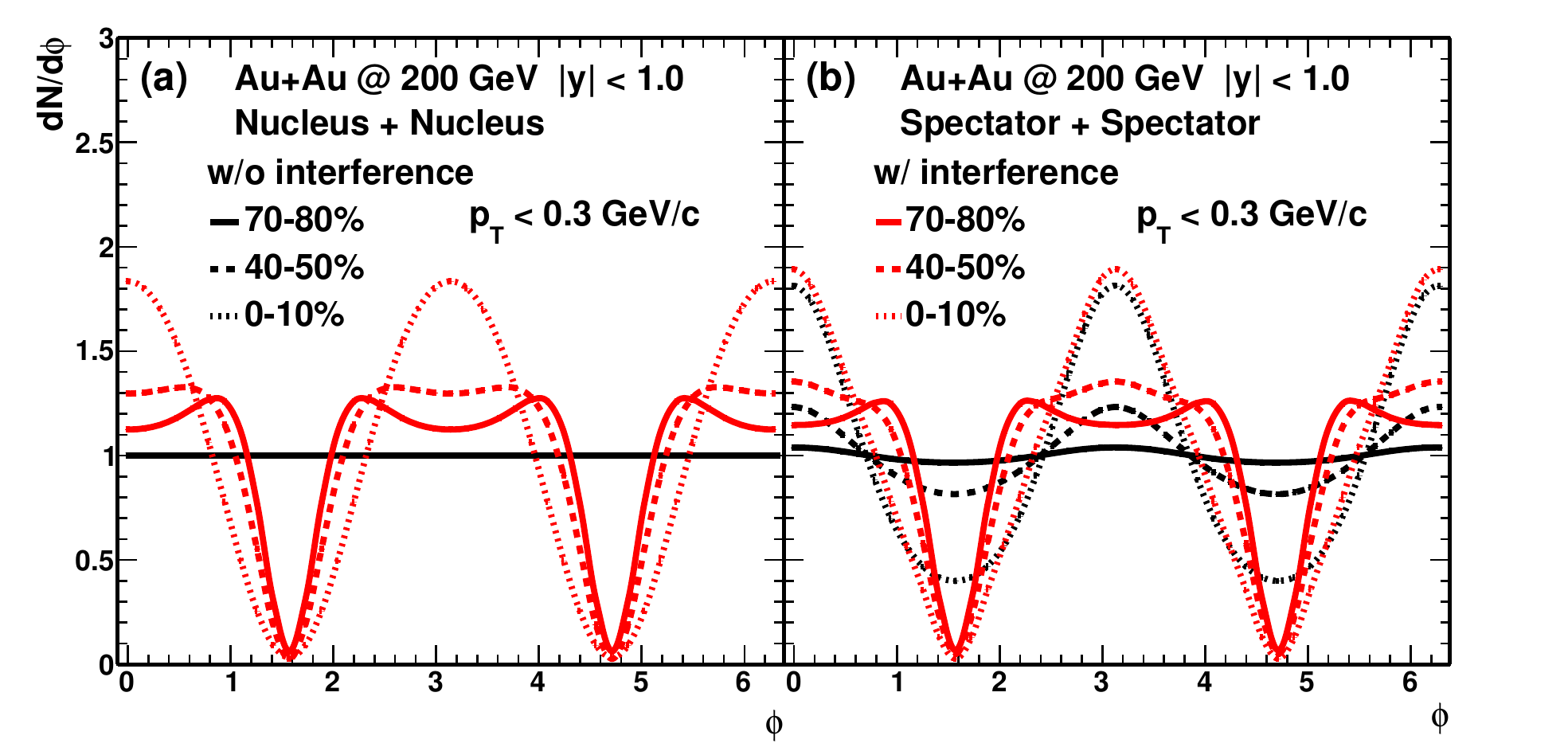}
\caption{The angular distributions of the coherent J/$\psi$ with respect to the reaction plane at mid-rapidity ($|y|<1$) in Au+Au collisions at $\sqrt{s_\text{NN}}$ = 200 GeV in the scenarios of ``N+N'' (a) and ``S+S'' (b). All the distributions are normalized such that $\langle \frac{dN}{d\phi} \rangle$ = 1. The black curves are the calculations without INT2N, while the red ones are with INT2N.}
\label{figure3}
\end{figure}

In single UPCs, there is no special azimuthal direction. However, in HHICs, the reaction plane~\cite{PhysRevC.58.1671}, spanned by the impact parameter and the beam axis, can be determined from the azimuthal anisotropies of produced particles due to the asymmetric collision geometry. Figure~\ref{figure3} shows the angular distributions of the coherent J/$\psi$ in the momentum space with respect to the reaction plane at mid-rapidity ($|y|<1$) in Au+Au collisions at $\sqrt{s_\text{NN}}$ = 200 GeV. The ``N+N'' and ``S+S'' scenarios are shown in panel(a) and (b), respectively. The black curves are the calculations without INT2N, while the red ones are with INT2N. Without INT2N effect, the coherent J/$\psi$ exhibits a uniform angular distribution in different centralities for the ``N+N'' scenario. However, in the ``S+S'' scenario, sizable anisotropy shows up due to the asymmetric density profile of the spectators, and the anisotropy grows larger towards more central collisions. When the INT2N is present, it drastically changes the angular distributions, leading to two dips at $\phi=\pi/2$ and $\phi=3\pi/2$ corresponding to the case where the $J/\psi$ $p_T$ is perpendicular to the reaction plane. The conventional anisotropy observed in HHICs arises from the anisotropy of the initial collision geometry that get preserved through strong parton-medium interactions. This anisotropy vanishes at low $p_T$ and in more central collisions, and is fundamentally different from the anisotropy seen for the coherent J/$\psi$, which originates from the asymmetric density profiles of the emitters convoluted with the INT2N effect. Hence, the measurement of J/$\psi$ angular distributions with respect to reaction plane in different centrality classes provides an additional handle to distinguish coherently produced J/$\psi$'s from ones produced in hadronic interactions. The resulting distributions for ``S+N'' and ``N+S'' scenarios are very close to those for the ``N+N'' and ``S+S'' scenarios, respectively. This indicates that the coherent J/$\psi$ anisotropy is also more sensitive to the Pomeron emitter, as for the case of $p_T$ spectrum.
\renewcommand{\floatpagefraction}{0.75}
\begin{figure}[htbp]
\includegraphics[keepaspectratio,width=0.5\textwidth]{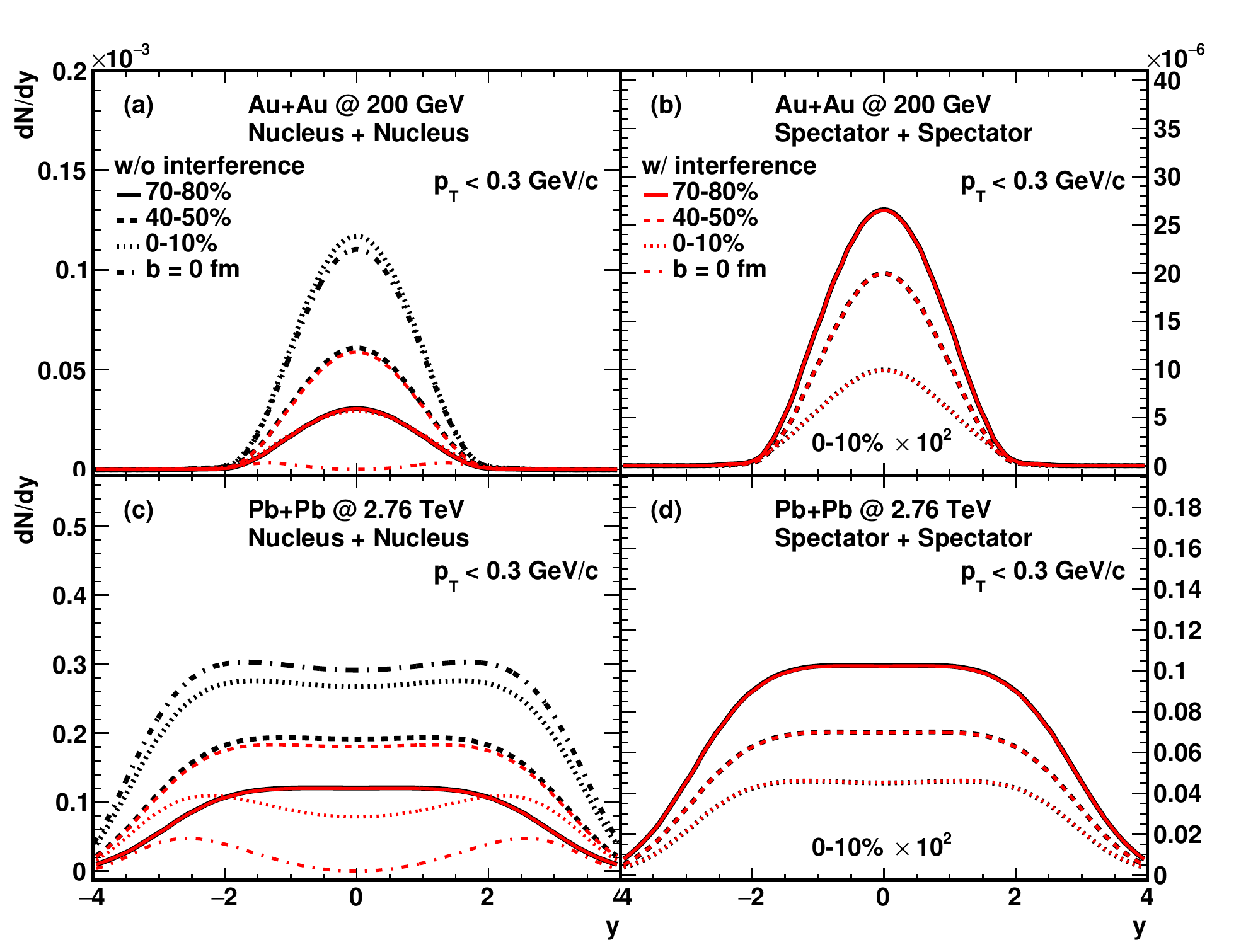}
\caption{The  J/$\psi$ $dN/dy$ distributions for different centralities with the scenarios of ``N+N'' (a and c) and ``S+S'' (b and d). The top two panels are for Au+Au collisions at $\sqrt{s_{\text{NN}}}$ = 200 GeV, while the bottom two show the calculations for Pb+Pb collisions at $\sqrt{s_{\text{NN}}}$ = 2.76 TeV. The black curves denote calculations without INT2N, while the red ones are with INT2N. The rapidity distributions for 0-10$\%$ central collisions in panels (b) and (d) are scaled by 10$^{2}$ for clarity. The curve for 0-10$\%$ central collisions with INT2N in panel (a) overlaps with the curves for 70-80$\%$ centrality.}
\label{figure6}
\end{figure}

For the total yield of coherent J/$\psi$, it is almost unaffected by the INT2N effect in UPCs, since the oscillation in the $\cos(\vec{p}_{T} \cdot \vec{b})$ term in Eq.~\ref{equation6} averages out as the J/$\psi$ $\langle p_T \rangle$ is significantly larger than $\langle \hbar/b \rangle$. In contrast, in HHICs, $\langle p_{T} \rangle \sim \langle \hbar/b \rangle$, and therefore the INT2N could significantly reduce the total cross-section, especially near mid-rapidity where the amplitudes for the two interference terms are similar. Figure~\ref{figure6} shows the expected J/$\psi$ $dN/dy$ for different centralities with the ``N+N'' (panels a and c) and ``S+S'' (panels b and d) scenarios. The top two panels are for Au+Au collisions at $\sqrt{s_{\text{NN}}}$ = 200 GeV, while the bottom two are for Pb+Pb collisions at $\sqrt{s_{\text{NN}}}$ = 2.76 TeV. The black curves are the calculations without INT2N, while the red ones show results with INT2N. The rapidity distributions with the ``S+S'' scenario for 0-10$\%$ central collisions are scaled by 10$^{2}$ for clarity. The J/$\psi$ $dN/dy$ in a certain centrality bin is related to the cross-section ($d\sigma/dy$) via the following equation:
  \begin{equation}
  \label{equation10}
  \frac{d\sigma}{dy}(\text{J}/\psi) = \int_{b_{min}}^{b_{max}} 2 \pi b \frac{dN}{dy}(\text{J}/\psi, b) db ,
  \end{equation}
where $b_{min}$ and $b_{max}$ are the minimum and maximum impact parameters for a given centrality bin, and $\frac{dN}{dy}(\text{J}/\psi, b)$ is the number of produced J/$\psi$ per unit rapidity in collisions of  impact parameter $b$. For the ``N+N'' scenario, the INT2N has little effect on the production yield in peripheral collisions, while it reduces the yield considerably in more central collisions. In particular, the coherent J/$\psi$ production is completely eliminated by the INT2N for the limiting case of $b = 0$ at $y = 0$. As shown in panel b and d, the coherent production of J/$\psi$ is almost unaffected by the INT2N in the ``S+S'' scenario due to the relatively large distance between the spectator nucleons. The resulting distributions for the ``S+N'' and ``N+S'' scenarios are very close to those of the ``N+N'' and ``S+S'' scenarios, respectively.

In summary, we have performed calculations of coherent J/$\psi$ photoproduction in HHICs with both the nucleus and spectator coupling hypotheses for photon and Pomeron emissions. In particular, the destructive interference in HHICs between photoproduction on ions moving in opposite directions is considered for the first time, which is found to significantly affect the coherent J/$\psi$ production. All four scenarios with the INT2N effect can describe the experimental data from ALICE within uncertainties. The difference in coherent J/$\psi$ production yields between different coupling assumptions is small for peripheral collisions and becomes significant in central collisions. Therefore, precise measurements towards central collisions are essential to distinguish the different scenarios. We have also studied the differential distributions for coherent J/$\psi$ as a function of transverse momentum, azimuthal angle and rapidity. All of them are found to be more sensitive to the Pomeron emitter rather than the photon emitter. Furthermore, these distributions are strongly modified by the INT2N effect, and can be confronted with future experimental measurements to test the presence of the INT2N effect. In present calculations, the nPDFs and possible hot medium effects are not considered yet, which can be included in future work.

%The current measurements in peripheral collisions can not rule out any of the scenarios. More precise measurements towards central collisions are essential to distinguish the different scenarios. The transverse momentum and angular distributions of J/$\psi$ production are also studied to test the different scenarios. The differential $p_{T}$ and $\phi$ distributions are sensitive to the Pomeron emitter. If the Pomeron couples with spectator, different $p_{T}$ spectra and $\phi$ distributions will be observed in different centralities, which may serve as a good probe to test the initial geometry in hadronic collisions. The effect of interference have also been discussed in the letter. The differential $p_{T}$ spectrum, angular distribution, and  total production has been dramatically changed by interference towards central collisions. Currently, the cold nuclear matter effects and possible hot medium effects are not considered, but they could be included in the future.

This work was funded by the National Natural Science Foundation of China under Grant Nos. 11505180 and 11375172, the U.S. DOE Office of Science under contract No. DE-AC-76SF00098 and DE-SC0012704, and MOST under Grant No. 2014CB845400.

\nocite{*}
\bibliographystyle{aipnum4-1}
\bibliography{aps}
\end{document}